\documentclass[aps, preprint, groupedaddress, floatfix]{revtex4}

\usepackage{epsfig}
\usepackage{epstopdf}
\usepackage{graphicx}
\usepackage{amsmath}
\usepackage{color}

\begin{document}

\title{Spin-density functional theories and their $+U$ and $+J$ extensions: 
a comparative study of transition metals and transition metal oxides}

\author{Hanghui Chen$^{1,2}$ and Andrew J. Millis$^{1}$}

\affiliation{
 $^1$Department of Physics, Columbia University, New York, NY, 10027, USA\\
 $^2$Department of Applied Physics and Applied Mathematics, Columbia University, New York, NY, 10027, USA\\}
\date{\today}

\begin{abstract}
Previous work on the physical content of exchange correlation
functionals that depend on both charge and spin densities is extended to
elemental transition metals and a wider range of perovskite transition
metal oxides. A comparison of spectra and magnetic moments calculated
using exchange correlation functionals depending on charge density
only or on both charge and spin densities, as well as the $+U$ and $+J$
extensions of these methods confirms previous conclusions that the
spin-dependent part of the exchange correlation functional provides an
effective Hund's interaction acting on the transition metal $d$
orbitals. For the local spin density approximation and spin-dependent
Perdew-Burke-Ernzerhof generalized gradient approximation, the
effective Hund's exchange is found to be larger than 1 eV. The results
indicate that at least as far as applications to transition metals and
their oxides are concerned, $+U$, $+J$ and +dynamical mean field
theory extensions of density functional theory should be based on
exchange-correlation functionals of charge density only.
\end{abstract}

\maketitle

\section{Introduction}

Density functional theory (DFT) is an enormously successful and
powerful method for treating the properties of interacting electrons
in atoms, molecules and solids~\cite{Jones89}. In its original form,
DFT was based on the minimization of a functional of the
space-dependent electronic charge density~\cite{Hohenberg-PR-1964},
but soon after, extensions to functionals depending on the spin
density as well as the charge density were
introduced~\cite{Kohn-PR-1965,Vonbarth19721629}. These functionals
are not exactly known, but current approximations to the charge-density-only 
functional such as the local density approximation
(LDA)~\cite{Kohn-PR-1965} and the generalized gradient approximation
(GGA)~\cite{Ma196818} provide a very good representation of the
electronic properties of many materials. Spin dependent extensions of
the local density approximation (LSDA)
~\cite{Vonbarth19721629,Rajagopal19731912,Gunnarsson19764274} and of
the generalized gradient approximation
(sGGA)~\cite{Rasolt19773234,Rasolt198145} provide important insights
into magnetic properties of many materials. However, the currently
available implementations of DFT have difficulty dealing with
phenomena associated with strong electronic correlations, including
magnetism and metal-insulator transitions \cite{Imada-RMP-1998},
associated with partly filled transition metal $d$-shells or partly
filled lanthanides $f$-shells. These
difficulties have motivated extensions of the original density
functional idea to explicitly include additional interaction terms
amongst physically relevant orbitals such as transition metal
$d$-orbitals~\cite{Anisimov-PRB-1991,Lie-PRB-1995}. Loosely speaking,
the extra interactions consist of a term, typically referred to as
``$U$'', that couples to the square of the total occupancy of the
selected orbitals and a set of terms, typically referred to as
``$J$'', that distinguish different multiplets at fixed total
occupancy of the $d$-shells. When the interaction effects are treated
within a Hartree-Fock approximation, the extensions are typically
referred to as ``$+U$'' and ``$+J$'' methods. When the interaction
physics is solved via the dynamical mean field method, the extension
~\cite{Georges04,Kotliar06,Held06} is referred to as ``+DMFT'' .

A key aspect of correlation physics in transition metals and their
oxides is the formation and dynamics of local moments arising from
electrons in partially filled transition metal $d$-shells. Both the
spin-dependent DFT (sDFT) methods and the $+U$/$+J$ extensions of DFT
express important aspects of this physics, and a combined sDFT+$U$+$J$
methodology seems an attractive approach to strong correlation
physics. However, recent studies indicate that this combination
produces seemingly unphysical behavior, including an unreasonable
$J$-dependence of structural parameters in
nickelates~\cite{Park:14,Park-PRB-2015} and of the high-spin/low-spin
energy difference in a spin crossover
molecule~\cite{ChenJia-PRB-2015}. A study by Park, Marianetti and 
one of us~\cite{Park-PRB-2015} on beyond-DFT theories for the rare earth
nickelates led to the conclusion that a source of the difficulty was
that the sDFT theories contain an effective $J$ acting on the Ni
$d$-states that is already larger than the value considered to be
reasonable for transition metals.

In this paper we extend the analysis of Ref.~\cite{Park-PRB-2015} to
wider classes of materials and additional observables. We study
SrMnO$_3$ (an antiferromagnetic insulator with a $d^3$ formal valence,
of current interest for potential multiferroic
behavior~\cite{Lee-PRL-2010, Rondinelli-PRB-2009}), SrVO$_3$ (a
moderately correlated metal with formal transition metal valence
$d^1$), and elemental Fe. For completeness we also present results for
the previously studied LaNiO$_3$. We restrict attention to a
Hartree-Fock treatment of the additional correlations (i.e. consider
only $+U$/$+J$ extensions but not +DMFT, although we expect our
conclusions will apply to that case also). We compute energy
differences between ferromagnetic and antiferromagnetic states as well
as magnetic moments. Further, we display the spin-dependent density of
states, which provides insight into the issues. Following
Ref.~\cite{Park-PRB-2015}, we compare results obtained from sDFT
theories to results of sDFT+$U$+$J$ and DFT+$U$+$J$ theories. We find
that DFT+$U$+$J$ with $J\sim$ 1-1.5 eV reproduces most aspects of
sDFT+$U$ ($J=0$) calculations, confirming that the conclusions of
Ref.~\cite{Park-PRB-2015} apply to a wide range of transition
metal-based materials. We show explicitly that in these systems, the
$+U$/$+J$ extensions of charge-density-only DFT provide a better
description of the physical properties than $+U$/$+J$ extensions of
sDFT.

The rest of this paper is organized as
follows. Sec.~\ref{sec:formalism} presents the formalism we
use. Sec.~\ref{sec:magnetization} presents energy differences between
different magnetic states and magnetic moments for ferromagnetic and
antiferromagnetic states. Sec.~\ref{sec:spectra} presents an analysis
of calculated densities of states. Sec.~\ref{sec:conclusion} is a
summary and conclusion.

\section{Formalism \label{sec:formalism}}

\subsection{Theoretical Approach}

Density functional theory (DFT) and spin-dependent density functional theory
(sDFT) and their $+U$ and $+J$ extensions are based on extremization of
functionals of charge density $n(\textbf{r})$, spin density
$m(\textbf{r})$ and the reduced density matrix describing the charge
$n_a$ and spin $m_a$ state of designated correlated orbitals labelled
by $a$. The extremization is actually accomplished by solving a
Schr\"{o}dinger equation involving an exchange-correlation potential
$V_{XC}$ which depends on $n(\textbf{r})$ (in the case of DFT) or on
$n(\textbf{r})$ and $m(\textbf{r})$ (in the case of sDFT) and an
additional functional that depends on the orbital occupancies and on
the interaction parameters (local-$d$ and intra-$d$ orbitals in the
usual applications to transition metals and their oxides):
schematically $V_U(n_a,m_a;U,J)$. An important part of the additional
functional is a double counting correction $V_{DC}$ that removes from
$V_{XC}$ the terms that are present in $V_U$. Thus in the ``DFT+''
methodologies $V_{DC}$ does not have spin dependence, whereas in the
``sDFT+'' methodologies it does.

The known exchange-correlation functionals depend on the full charge
(spin) density, the portion pertaining to the designated correlated
orbitals cannot be extracted and the double counting correction thus
cannot be rigorously derived~\cite{Haule-PRL-2015}. The double
counting term must be specified by approximate, phenomenologically
based arguments. Different forms have been introduced
~\cite{Czyzyk-PRB-1994, Karolak10}. In this study, we use the
widely-adopted fully localized limit (FLL) form. However, our basic
conclusions are independent of the precise form chosen. 

For the case of DFT+$U$+$J$, 
the FLL double counting correction reads:
\begin{equation}
\label{eqn:dc1} V_{DC} = U\left(N_d-\frac{1}{2}\right)-J
\left(\frac{1}{2}N_{d}-\frac{1}{2}\right)
\end{equation}
where $N_d$ is the total occupancy of designated correlated
orbitals (here transition metal $d$ orbitals). $U$ is the 
Hubbard $U$ and $J$ is the Hund's coupling, which are
the standard inputs of DFT+$U$+$J$ calculations.

For the case of sDFT+$U$+$J$, $V_{DC}$ is spin dependent and the explicit 
FLL double counting form reads:

\begin{equation}
\label{eqn:dc2} V^{\sigma}_{DC} = U\left(N_d-\frac{1}{2}\right)-J
\left(N^{\sigma}_{d}-\frac{1}{2}\right)
\end{equation}
where $N^{\sigma}_d$ is the total occupancy of designated orbitals
with spin $\sigma$. $N_d=\sum_{\sigma}N^{\sigma}_d$. $U$
and $J$ have the same meaning as in Eq.~(\ref{eqn:dc1}). For
non-magnetic cases, $N^{\sigma}_d = \frac{1}{2}N_d$ and
Eq.~(\ref{eqn:dc2}) reduces to Eq.~(\ref{eqn:dc1}).

In our studies we compare two forms of $V_{XC}(n(\textbf{r}))$: the
local density approximation (LDA)~\cite{Kohn-PR-1965} and the
generalized gradient approximation (GGA) with the
Perdew-Burke-Ernzerhof (PBE)
parametrization~\cite{Perdew-PRL-1996}. Correspondingly for the
spin-dependent density functionals, we use the local spin density
approximation (LSDA)~\cite{Kohn-PR-1965} and the spin-dependent GGA
with the PBE parametrization (sPBE)~\cite{Perdew-PRL-1996}. For the
$+U$ and $+J$ extensions, we use the rotationally invariant Hubbard/Hund's
corrections introduced by Liechtenstein \textit{et. al.}~\cite{Lie-PRB-1995}.

We note that DFT+$U$+$J$ and sDFT+$U$+$J$ methods become equivalent if
applied to non-magnetic states ($m(\textbf{r}) = m_a = 0$). For
magnetic materials, the two methods differ
in principle because in the DFT+$U$+$J$ case only the spin-dependence
of the correlated orbitals (here transition metal $d$ orbitals)
contributes to the spin dependence of the self-consistent potential
felt by electrons. This is because the exchange-correlation potential
depends only on the total charge density, so it yields a spin-independent
contribution to the potential. In contrast, in the sDFT+$U$+$J$ case the
spin-dependence of the exchange-correlation potential means that the
spin polarization of the non-$d$ orbitals also contributes to the
spin-dependence of the self-consistent potential. However we shall see
that for the situations we consider, this difference is unimportant in
practice, probably because the polarization of the non-correlated
orbitals is small. The key difference between different choices of
exchange-correlation functionals will be seen to be the magnitude of the
spin-dependent term acting on the correlated orbitals.

\subsection{Computational Details}

\begin{figure}[t]
\includegraphics[angle=0, width=0.9\columnwidth]{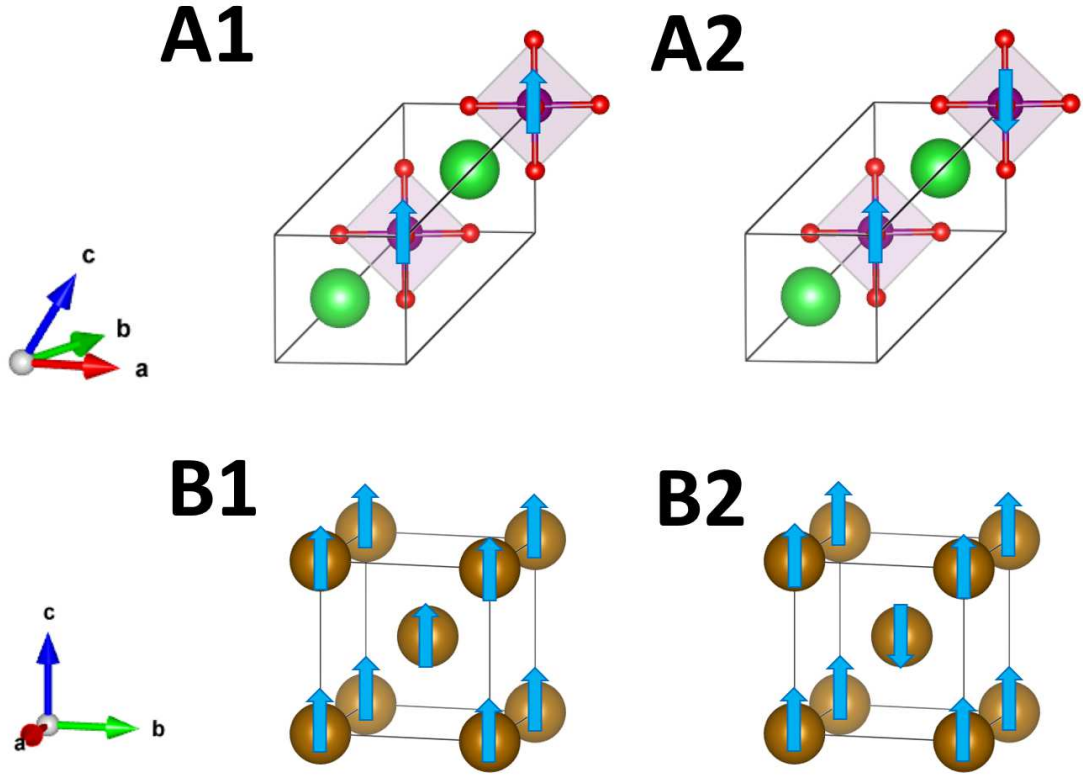}
\caption{\label{fig:cell} Computational unit cells showing atoms
  (balls) and spin alignments (arrows). Panel \textbf{A}: simulation
  cell for transition metal oxides $AM$O$_3$. The two perovskite unit
  cells are stacked along the [111] direction. The $A$-site ion ($A$ =
  La or Sr in the current study) is the large ball (green on-line) and
  the intermediate-sized ball (purple on-line) represents the
  transition metal ($M$) ion ($M$ = Mn, V or Ni in the current
  study). The small balls (red on-line) represent oxygen atoms. Panel
  \textbf{B}: simulation cell for body-centered iron.  Column
  \textbf{1}: ferromagnetic ordering and column \textbf{2}: $G$-type
  antiferromagnetic ordering. }
\end{figure}

We present results for representative transition metal oxides: cubic
SrMnO$_3$, cubic SrVO$_3$ and pseudo-cubic LaNiO$_3$ (the last
compound was previously studied in Ref.~\cite{Park-PRB-2015} and we
reproduce the results for comparison) and one representative
transition metal: iron. The simulation cell is illustrated in
Fig.~\ref{fig:cell}. For transition metal oxides, it consists of two
perovskite primitive cells (10 atoms in total) stacked along the [111]
direction (panels \textbf{A} of Fig.~\ref{fig:cell}). For transition
metal, we study body-centered iron (panels \textbf{B} of
Fig.~\ref{fig:cell}). For both transition metal oxides and transition
metals, the computational cell can accommodate both ferromagnetic
ordering and $G$-type (two-sublattice N\'{e}el) antiferromagnetic
ordering. We use experimental lattice constants, respectively
3.80~\AA~(SrMnO$_3$~\cite{Sondena-PRB-2007}),
3.84~\AA~(SrVO$_3$~\cite{Maekawa-JAC-2006}) and
3.86~\AA~(LaNiO$_3$~\cite{Kwang-JKPS-1999}) and
2.86~\AA~(Fe~\cite{Davey-PR-1925}).

The density functional theory
calculations~\cite{Hohenberg-PR-1964,Kohn-PR-1965} are performed
within the~\textit{ab initio} plane-wave approach
~\cite{Payne-RMP-1992}, as implemented in the Vienna Ab-initio
Simulation Package (VASP)~\cite{Kresse-PRB-1996}.  We employ projector
augmented wave (PAW)
pseudopotentials~\cite{Blochl-PRB-1994,Kresse-PRB-1999}.  We use an
energy cutoff 600 eV and a $10\times 10\times 10$ Monkhorst-Pack
grid. A higher energy cutoff (800 eV) and a denser 
$k$-point sampling ($12\times12\times12$) are used to test the 
convergence and no significant difference is found.
All the calculations allow for the possibility of
spin-polarization to study different types of long-range magnetic
orderings (if they can be stabilized). 
LDA+$U$+$J$ and PBE+$U$+$J$ are implemented in VASP as
LDAUTYPE=4 and LSDA+$U$+$J$ and sPBE+$U$+$J$ are implemented in
VASP as LDAUTYPE=1. 

\begin{figure}[t]
\includegraphics[angle=0, width=0.9\columnwidth]{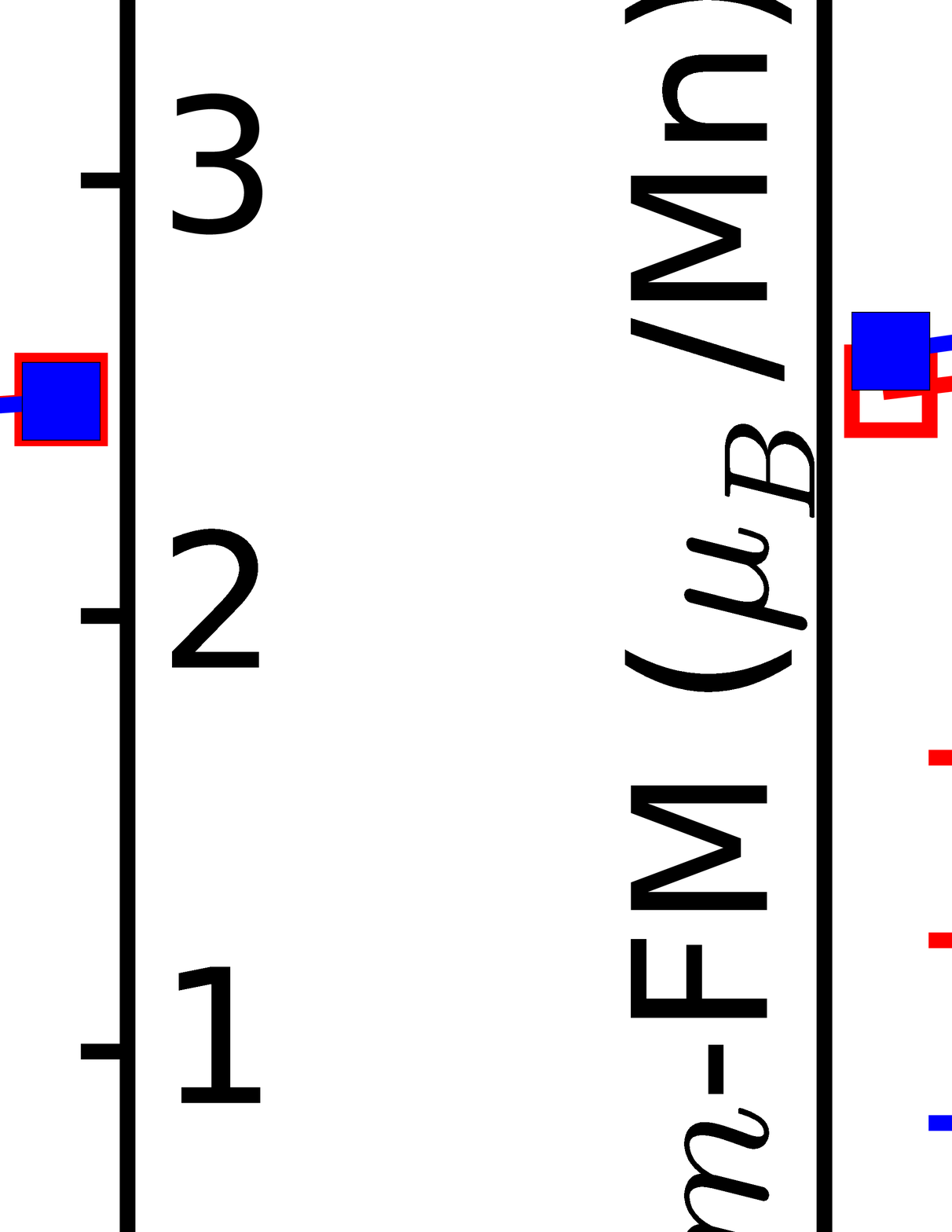}
\caption{\label{fig:compare} Comparison of predictions from 
  LDA+$U$+$J$/PBE+$U$+$J$ (panels~\textbf{A}) and
  LSDA+$U$+$J$/sPBE+$U$+$J$ (panels~\textbf{B}) methods for ground
  state properties of cubic SrMnO$_3$. Row~\textbf{1}: energy
  difference $\Delta E = E(G) - E(F)$ between ferromagnetic ($F$) and
  $G$-type antiferromagnetic ($G$) ordering.  Row~\textbf{2}: magnetic
  moment per Mn of ferromagnetic state. Row~\textbf{3}: magnetic
  moment per Mn of $G$-type antiferromagnetic state.}
\end{figure}

\section{Magnetization and Energy Differences\label{sec:magnetization}}

In this section, we consider the ferromagnetic-antiferromagnetic
energy differences and local magnetic moments in the ferromagnetic and
antiferromagnetic states obtained using different methods.  We begin
with SrMnO$_3$, a cubic perovskite antiferromagnetic insulator known
experimentally ~\cite{Sondena-PRB-2007} to exhibit an approximately a
high-spin $d^3$ configuration with a fully spin-polarized $t_{2g}$
shell and a nearly empty $e_g$ shell.

Panel~\textbf{A1} of Fig.~\ref{fig:compare} presents the energy
difference between $G$-type (two sublattice N\'{e}el)
antiferromagnetic and ferromagnetic states calculated using the
DFT+$U$+$J$ method with two choices of exchange-correlation potential:
the local density approximation (LDA) and the generalized gradient
approximation in the Perdew-Burke-Ernzerhof parameterization
(PBE). Panels~\textbf{A2, A3} present the local magnetic moments of the
ferromagnetic and antiferromagnetic states, respectively.  We require
that the net on-site interaction is repulsive: this imposes the
constraint that $U > 3J$. Therefore, for $J=$ 1 eV, we only consider
$U > 3$ eV. In the DFT+$U$+$J$ method, there is no intrinsic exchange
splitting in the exchange correlation functionals.

We see immediately that the two density functionals, LDA and PBE, give
essentially identical results. For pure LDA and PBE ($U=J=0$), SrMnO$_3$ is
predicted to be non-magnetic. For moderate $U=2$ eV, the ground state is
antiferromagnetic and a ferromagnetic state could not be
stabilized. For larger $U\gtrsim 4$ eV, the ferromagnetic state is
locally stable. For sufficiently large $U$, the ground state is
ferromagnetic. Increasing $J$ favors ferromagnetism. For $U < 4$ eV,
the calculated moments are substantially below the experimental value
of 2.6 $\mu_B$/Mn~\cite{Takeda-JPSJ-1974}. We therefore believe that to
adequately represent the physics of SrMnO$_3$ within the DFT+$U$+$J$
method a $U\gtrsim 4$ eV is required. For sufficiently large $U$ and
$J$ ($J\gtrsim 1$ eV for $U = 6$ eV or $J\gtrsim 0.8$ eV for $U = 8$
eV), the calculated ground state of SrMnO$_3$ is ferromagnetic instead
of experimentally observed $G$-type antiferromagnetic.  We therefore
believe that $U \lesssim 8$ eV is required within this method.

Magnetism arises from a Hartree treatment of the $U$ interaction,
supplemented by the tendency of the $J$ term to favor high-spin
states.  As $U$ is increased above $U = 4$ eV or $J$ is increased from
$J = 0$, the energy of the ferromagnetic state decreases relative to
that of the antiferromagnetic state. The change in energy of the two
states can be explained in terms of the energy dependence of the
relevant exchange processes. Antiferromagnetism results from an
inter-$t_{2g}$ superexchange $\sim t^2/(U+3J)$ where $t$ is the hopping, 
while ferromagnetism comes from double exchange mediated by virtual 
occupancy of the $e_g$ and proportional to $t$. As $U$ and $J$ are increased, 
the antiferromagnetic interaction thus weakens and above some critical 
$U_c$ and $J_c$, the ferromagnetic interaction dominates.
 
We next consider the predictions of the spin dependent density
functionals, shown in the panels \textbf{B} of Fig.~\ref{fig:compare}.
We first observe that LSDA+$U$+$J$ and sPBE+$U$+$J$ produce different
results, with sPBE+$U$+$J$ favoring ferromagnetism more than
LSDA+$U$+$J$ and predicting slightly larger moments. Even without the
$+U$/$+J$ corrections, pure LSDA and sPBE stabilize both ferromagnetic 
and antiferromagnetic states with local magnetic moments
close to the experimental values.  We interpret this
result as indicating that the spin-dependent functionals possess an
intrinsic exchange splitting that is large enough to separate the
lower and upper Hubbard bands of Mn-$d$ states, consistent with
previous findings of Ref.~\cite{Park-PRB-2015} in the context
of rare earth nickelates. We also comment that it is widely
known~\cite{Luo-PRL-2007,Burton-PRB-2009, Chen-PRB-2012,
Chen-Nano-2014} that sPBE gives a reasonable description of
magnetic properties of La$_{1-x}$Sr$_x$MnO$_3$ (in particular, the
magnetic transition point around $x$ = 0.5), while adding $U$ to
sPBE impairs the agreement between theory and experiment. However,
on the other hand, the physical $U$ on Mn $d$-orbitals is
definitely nonzero (around 4 eV from constrained random phase
approximation calculations, cRPA~\cite{Vaugier-PRB-2012}). Our
results provide a natural explanation that the intrinsic ``$J$" in
the sPBE already produces a large enough spin-splitting and 
adding $U$ further splits spin channels, which thus leads to some unphysical
results. Using LDA+$U$+$J$/PBE+$U$+$J$, we find that a physical
range of $U$ is between 4 and 8 eV, which is more consistent with
previous cRPA calculations.

As was found in DFT+$U$+$J$ calculations, increasing $U$ in
sDFT+$U$+$J$ decreases the energy difference between the
antiferromagnetic and ferromagnetic states, so that for large enough
$U$ the ferromagnetic state becomes favored. However, in contrast to
DFT+$U$+$J$, increasing $J$ in sDFT+$U$+$J$ \textit{destabilizes} the 
ferromagnetic state. This counterintuitive result is similar to
the previous finding of $J$-dependence of the
high-spin/low-spin transition point in a spin crossover
molecule~\cite{ChenJia-PRB-2015} and is discussed in more
detail in the next section.

\begin{figure}[t]
\includegraphics[angle=0, width=0.95\textwidth]{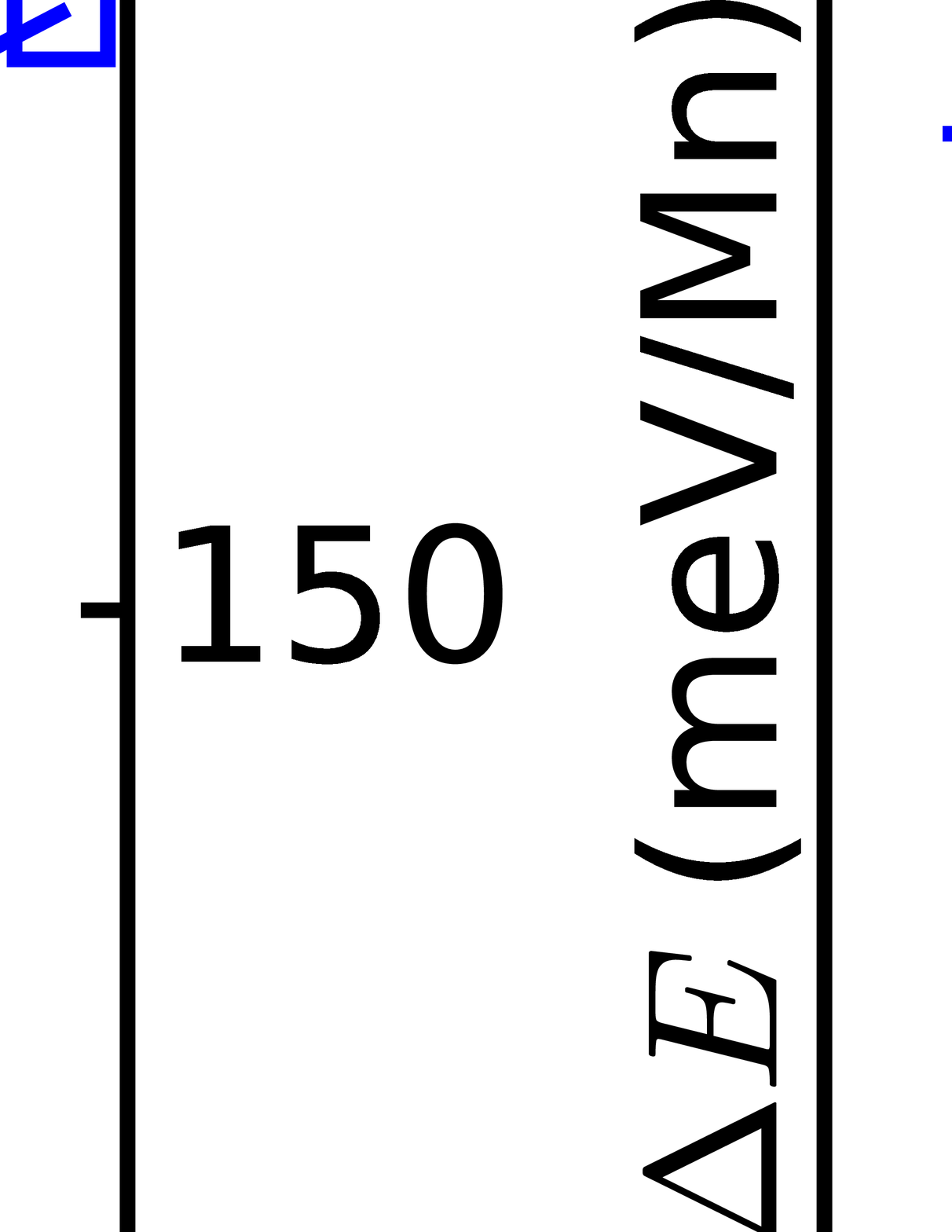}
\caption{\label{fig:relationLDA} Comparison of
  antiferromagnetic-ferromagnetic ground state energy differences
  obtained using sDFT+$U$ (closed symbols, red on-line) and
  DFT+$U$+$J$ with $J$ = 1 eV (open symbols, blue on-line) for
  materials indicated.  Left panels: LSDA+$U$ and LDA+$U$+$J$ (with
  $J$ = 1 eV).  Right panels: sPBE+$U$ and PBE+$U$+$J$ (with $J$ = 1
  eV).}
\end{figure}

\begin{figure}[t]
\includegraphics[angle=0, width=0.95\textwidth]{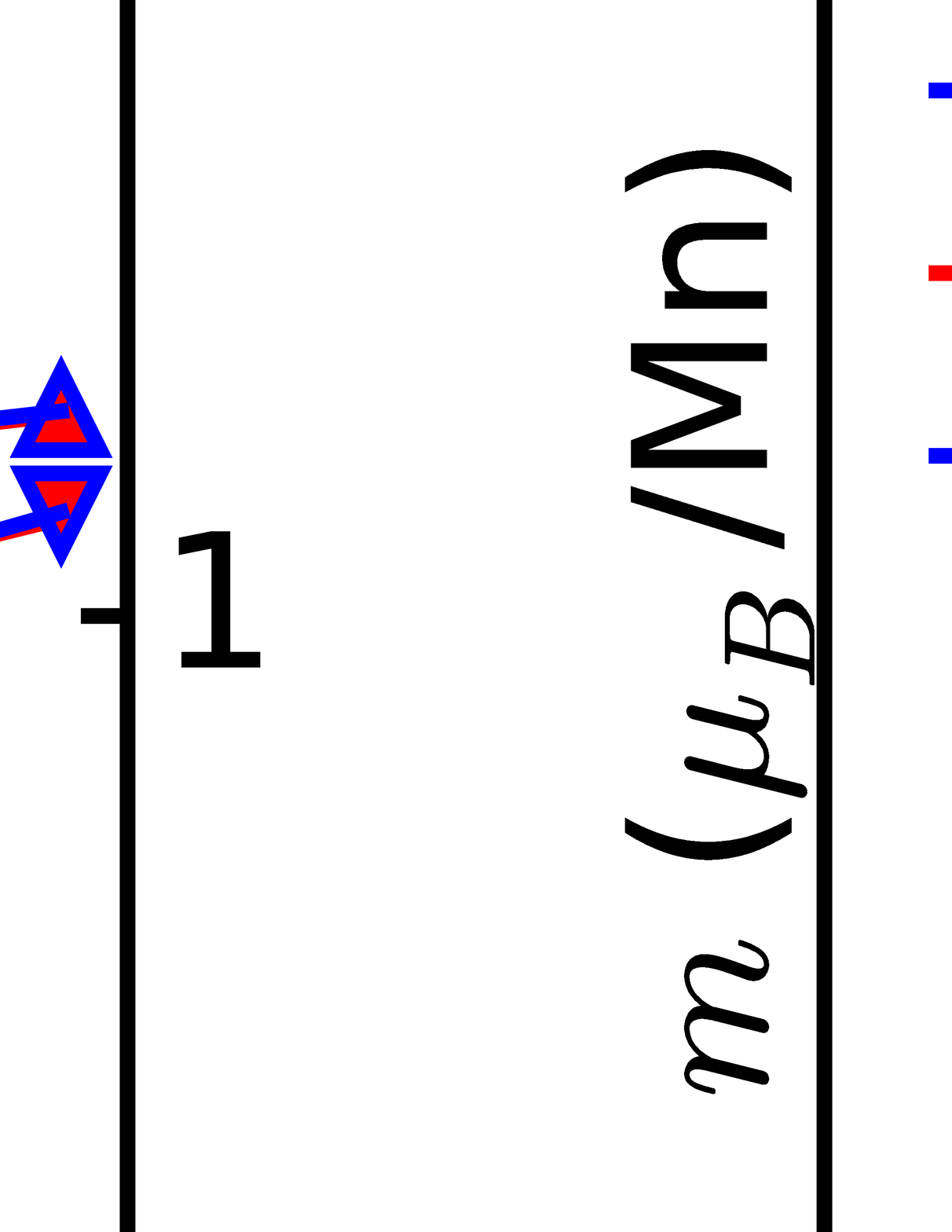}
\caption{\label{fig:relationPBE} Comparison of magnetic moments
  obtained using sDFT+$U$ (closed symbols, red on-line) and
  DFT+$U$+$J$ with $J$ = 1 eV (open symbols, blue on-line) 
  for materials indicated.
  Left panels: LSDA+$U$ and LDA+$U$+$J$ (with $J$ = 1 eV).
  Right panels: sPBE+$U$ and PBE+$U$+$J$ (with $J$ = 1 eV).
  The upper triangles are for ferromagnetism.  The down triangles are for
  $G$-type antiferromagnetism.}
\end{figure}

To further investigate the differences between DFT+$U$+$J$ and
sDFT+$U$+$J$ methods and to understand the
robustness of our results across the perovskite family of materials,
we present in Fig.~\ref{fig:relationLDA} the
ferromagnetic-antiferromagnetic energy difference $E(G) - E(F)$ of
different transition metal oxides, calculated using sDFT+$U$ (with $J$
= 0) and DFT+$U$+$J$ (with $J$ = 1 eV).  We compare SrMnO$_3$
(antiferromagnetic insulator with half-filled $t_{2g}$-shell),
SrVO$_3$ (moderately correlated metal) and LaNiO$_3$ (negative
charge-transfer metal).

Fig.~\ref{fig:relationLDA} shows clearly that increasing $J$ brings
the DFT$+U+J$ results into closer agreement with the results of sDFT$+U$
($J$ = 0) calculations, indicating that in transition metal perovskites
the main physical content of the spin-dependent density functionals is
an effective ``$J$'' acting on the transition metal $d$-levels.  We may
define the size of the effective ``$J$'' of the sDFT functionals as the
$J$ that needs to be added to make the DFT$+U+J$ results 
coincide with the sDFT+$U$ ($J$ = 0) results. The effective $J$ is $\gtrsim
1$ eV and is seen to depend on materials and functionals, being larger 
for sPBE than for sDFT and larger for LaNiO$_3$ than for SrMnO$_3$.

Fig.~\ref{fig:relationPBE} shows the magnetic moments of different
transition metal oxides, calculated as in Fig.~\ref{fig:relationLDA}
and presented using the same conventions. Consistent with
Fig.~\ref{fig:relationLDA}, a $J$ equal to or slightly larger than 1
eV must be added in the spin-independent DFT+$U$+$J$ calculations to
reproduce the magnetic moments calculated from the sDFT+$U$ ($J$
= 0) method.

\begin{figure}[t]
\includegraphics[angle=0, width=0.95\textwidth]{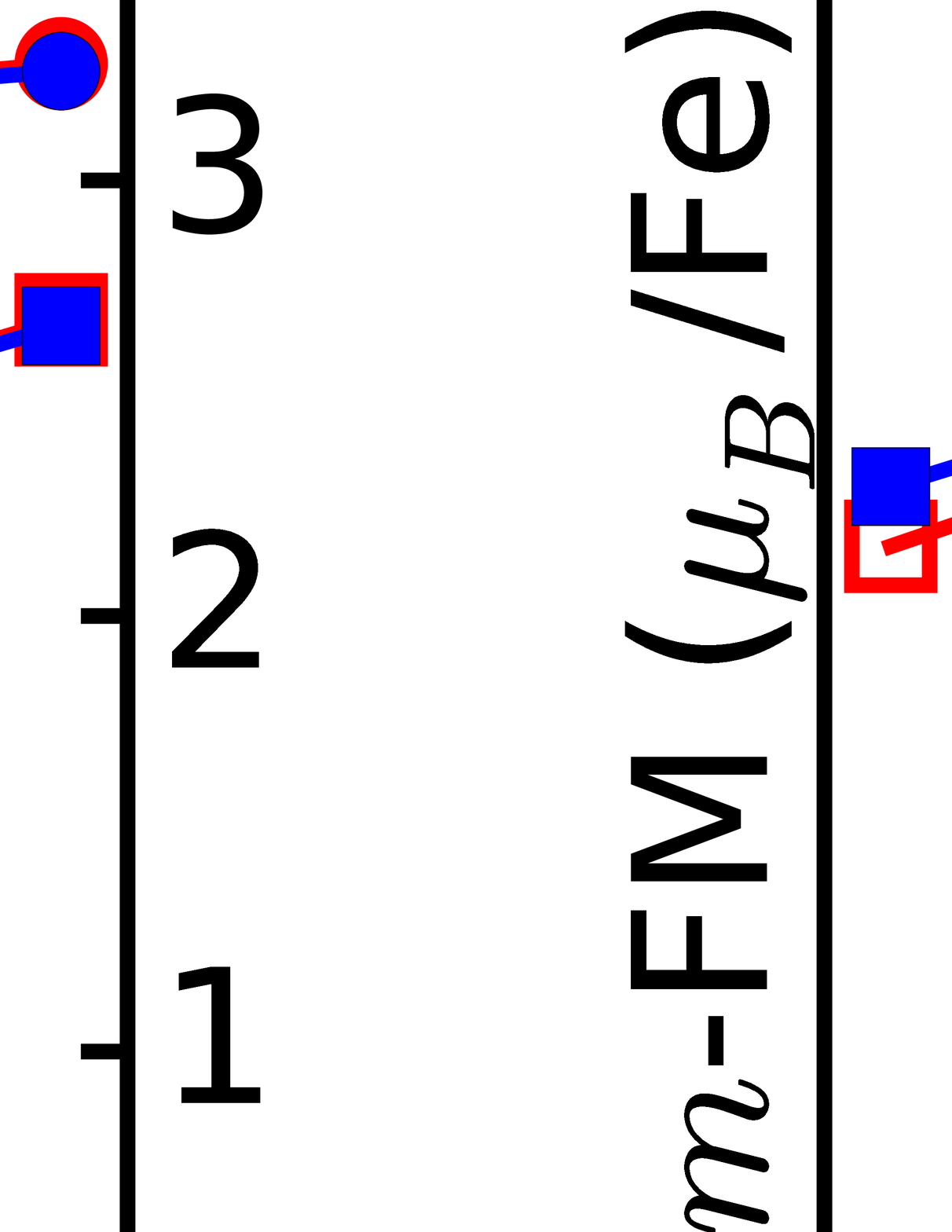}
\caption{\label{fig:compare_Fe} Comparison of predictions from
  LDA+$U$+$J$/PBE+$U$+$J$ (panels~\textbf{A}) and
  LSDA+$U$+$J$/sPBE+$U$+$J$ (panels~\textbf{B}) methods for ground
  state properties of body-centered Fe. Row~\textbf{1}: energy
  difference $\Delta E = E(G) - E(F)$ between ferromagnetic ($F$) and
  $G$-type antiferromagnetic ($G$) ordering.  Row~\textbf{2}: magnetic
  moment per Mn of ferromagnetic state. Row~\textbf{3}: magnetic
  moment per Mn of $G$-type antiferromagnetic state.}
\end{figure}

We next consider Fe, which we study as a representative elemental
transition metal. We investigate the extent to which the previous
results we obtain from perovskite oxides may apply to transition
metals. Fig.~\ref{fig:compare_Fe}, which uses the same convention as
Fig.~\ref{fig:compare}, presents the antiferromagnetic-ferromagnetic
energy difference as well as the local moments in ferromagnetic and
antiferromagnetic states for body-centered Fe.  Experimentally, the
body-centered iron is ferromagnetic with a magnetic moment of 2.2
$\mu_B$/Fe~\cite{Crangle-PRSL-1963}. The left panels of
Fig.~\ref{fig:compare_Fe} show that a Hubbard $U$ less than 4 eV in
the DFT+$U$+$J$ method does not produce a magnetic ground state for
iron, which is inconsistent with experiment. As $U\geq$ 4 eV, a
magnetic ground state is produced with a sizable magnetic moment on Fe
($> 2 \mu_B$/Fe). As $J=0$, the ferromagnetic and antiferromagnetic
states are almost degenerate for a wide range of Hubbard $U$.  As $J$
is increased from $J = 0$, the ferromagenetic state becomes
substantially favored in energy.  Similar results were also found for
SrMnO$_3$ and other perovskite oxides. The right panels of
Fig.~\ref{fig:compare_Fe} show (also as found in perovskite oxides)
that LSDA/sPBE alone ($U$ = $J$ = 0) suffices to split the spin and
yield a sizable magnetic moment ($\sim$2.2 $\mu_B$/Fe for
ferromagnetism and $\sim$1.5 $\mu_B$/Fe for antiferromagnetism), which
agrees well with the experiment~\cite{Crangle-PRSL-1963}. Increasing
$U$ impairs the agreement and increasing $J$ in sDFT+$U$+$J$
destabilizes ferromagnetism.
 
\section{Density of States \label{sec:spectra}}

In this section, we study the density of states (DOS) obtained using
different exchange correlation functionals at $U$ = 0 and
6 eV and Hund's coupling $J$ = 0 and 1 eV. For ease of
interpretation, we present results obtained in the ferromagnetic state.
It is useful to analyze the results in terms of the standard
phenomenological Slater-Kanamori interaction, which for simplicity we
discuss for the simple case of a half-filled fully spin-polarized
orbitally symmetric $t_{2g}$ shell treated in the Hartree-Fock
(``+$U$+$J$'') approximation (this is a simple model for cubic SrMnO$_3$). 
In this case the spin up/down potential for each $t_{2g}$ orbital arising 
from this interaction is:
\begin{eqnarray}
V^{\uparrow}_{\alpha} = 2U-6J\label{Vup}\\
V^{\downarrow}_{\alpha} = 3U-4J\label{Vdown}
\end{eqnarray}
where $\alpha$ labels a $t_{2g}$ orbital (the derivation is in the Appendix).
Taking into account the double counting terms, in the DFT+$U$+$J$ method,
we have:
\begin{eqnarray}
V^{\uparrow}_{\alpha} = 2U-6J - V_{DC}\label{VupDFT}\\
V^{\downarrow}_{\alpha} = 3U-4J - V_{DC}\label{VdownDFT}
\end{eqnarray}
where the double counting correction $V_{DC}$ is spin-independent. Therefore, 
in the DFT+$U$+$J$ case the energy difference between the spin up
and spin down potentials is:

\begin{equation}
\label{eqn:DCDFT} |V^{\uparrow}_{\alpha} - V^{\downarrow}_{\alpha}| = U+2J 
\end{equation}
However, in the sDFT+$U$+$J$ method, we have:
\begin{eqnarray}
V^{\uparrow}_{\alpha} = 2U-6J - V^{\uparrow}_{DC}\label{VupSDFT}\\
V^{\downarrow}_{\alpha} = 3U-4J - V^{\downarrow}_{DC}\label{VdownSDFT}
\end{eqnarray}
where the double counting correction $V^{\sigma}_{DC}$ is spin
dependent. Therefore, in the sDFT+$U$+$J$ case, using the FLL double
counting scheme Eq.~(\ref{eqn:dc2}), the energy difference between the 
spin up and spin down potentials is:

\begin{equation}
\label{eqn:DCsDFT} |V^{\uparrow}_{\alpha} - V^{\downarrow}_{\alpha}| = |U+2J-Jm| 
= |U-J| 
\end{equation}
where $m = N^{\uparrow}_d -N^{\downarrow}_d$ is the magnetization of
the $d$ orbitals and in this simple $t_{2g}$ model $m=3$.  This shows
that adding $J$ in sDFT+$U$+$J$ \textit{reduces} the spin splitting.

\begin{figure}[t]
\includegraphics[angle=0, width=0.95\textwidth]{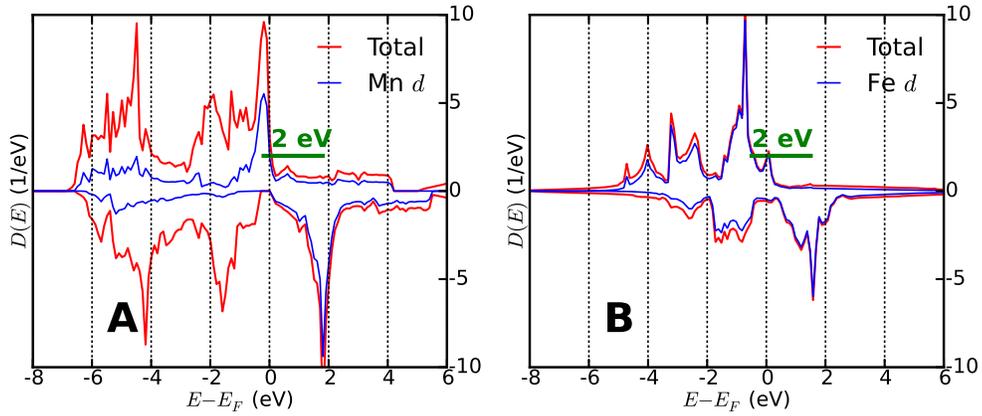}
\caption{\label{fig:smofedos} Density of states of ferromagnetic
  ordering calculated using the LSDA ($U$=$J$=0) method. \textbf{A})
  cubic SrMnO$_3$; \textbf{B}) iron. Positive and negative $y$ axis
  curves show majority and minority density of states, respectively.
  The horizontal green line and the number provide estimates of the
  spin splitting.}
\end{figure}

Next we present the DFT-computed densities of states in which we find the
peaks that are attributable to the $d$-levels. The energy differences
between the majority and minority spin channels then reflect the
values of $U$ and $J$. Fig.~\ref{fig:smofedos} presents the pure LSDA
densities of states for ferromagnetic SrMnO$_3$ and elemental Fe. We see
that both materials exhibit a DOS peak at an energy $\sim 2$ eV above
the Fermi level in the minority spin channel and a $d$-related peak in
the majority spin channel slightly below the Fermi level. In SrMnO$_3$
the $d$-states visible at energies $\sim -4$ to $-6$ eV arise from
admixture with oxygen orbitals. We define the spin splitting as the
peak-to-peak energy difference between the majority and minority spin
$d$-contributions to the densities of states and indicate it by the
heavy green line. For SrMnO$_3$ the peak to peak splitting of the
$d$-bands provides an estimate of intrinsic ``$2J$'', indicating an effective
$J$ of about 1 eV. In Fe the interpretation is complicated by the
higher occupancy of the $d$-band (the calculated $N_d$ of Fe is very
close to the nominal occupancy of 6).

\begin{figure}[t]
\includegraphics[angle=0, width=0.95\textwidth]{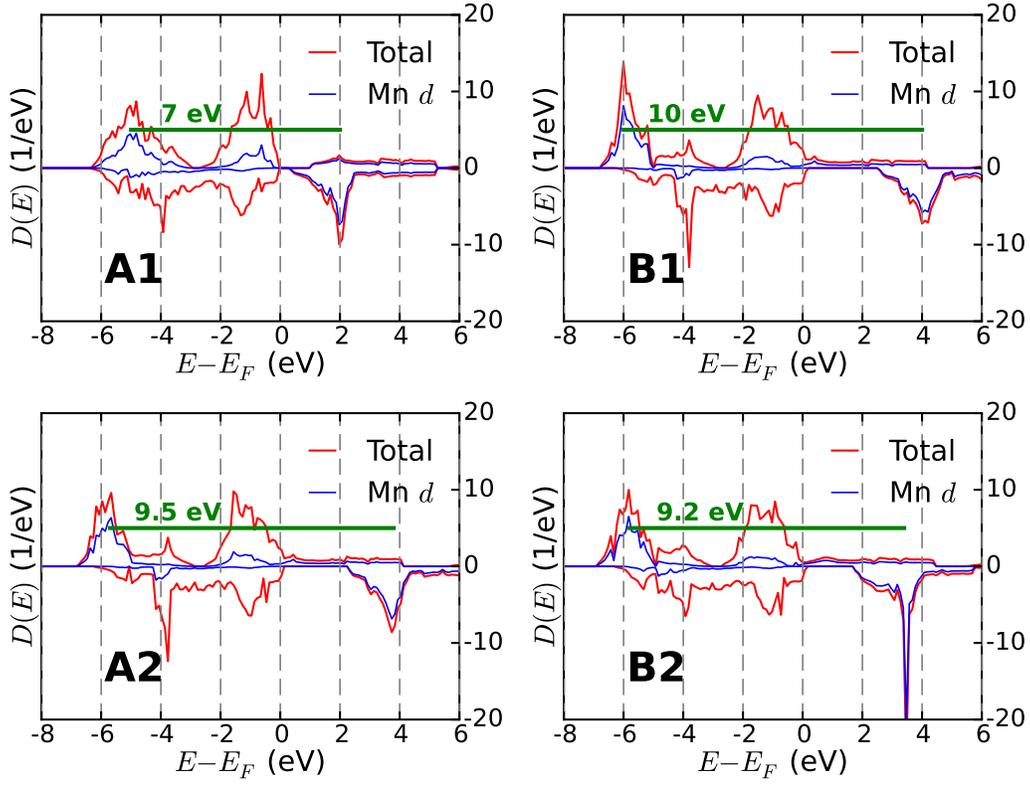}
\caption{\label{fig:dos} Total density of states (heavy line, red
  on-line) and Mn $d$-projected (light line, blue on-line) of cubic
  SrMnO$_3$ in the ferromagnetic state. Positive and negative $y$ axis
  curves show the majority and minority densities of states, respectively.
  \textbf{A1}) LDA+$U$; \textbf{A2}) LDA+$U$+$J$; \textbf{B1})
  LSDA+$U$; \textbf{B2}) LSDA+$U$+$J$; $U$ = 6 eV and $J$ = 1 eV.
  Horizontal green lines and numbers provide estimates of spin
  splitting obtained from peak-to-peak separation of majority and
  minority spin $d$-density of states peaks. }
\end{figure}

Fig.~\ref{fig:dos} presents the density of states for ferromagnetic
SrMnO$_3$ calculated using different exchange correlation functionals.
The majority spin density of states has a significant peak between -2
and 0 eV, but this peak has only modest $d$ content. It arises from
oxygen $p$ states, with modest $p$-$d$ hybridization. The main portion
of the occupied majority spin $d$-states occurs much further below the
Fermi level, at an energy of -5 to -6 eV with the precise energy
depending on the exchange correlation functional. In mathematical
terms the double counting correction shifts the mean energy of the
$d$-states down to this low energy (a level repulsion due to
hybridization with the oxygen $p$ states also plays a role).

\begin{figure}[t]
\includegraphics[angle=0, width=0.95\textwidth]{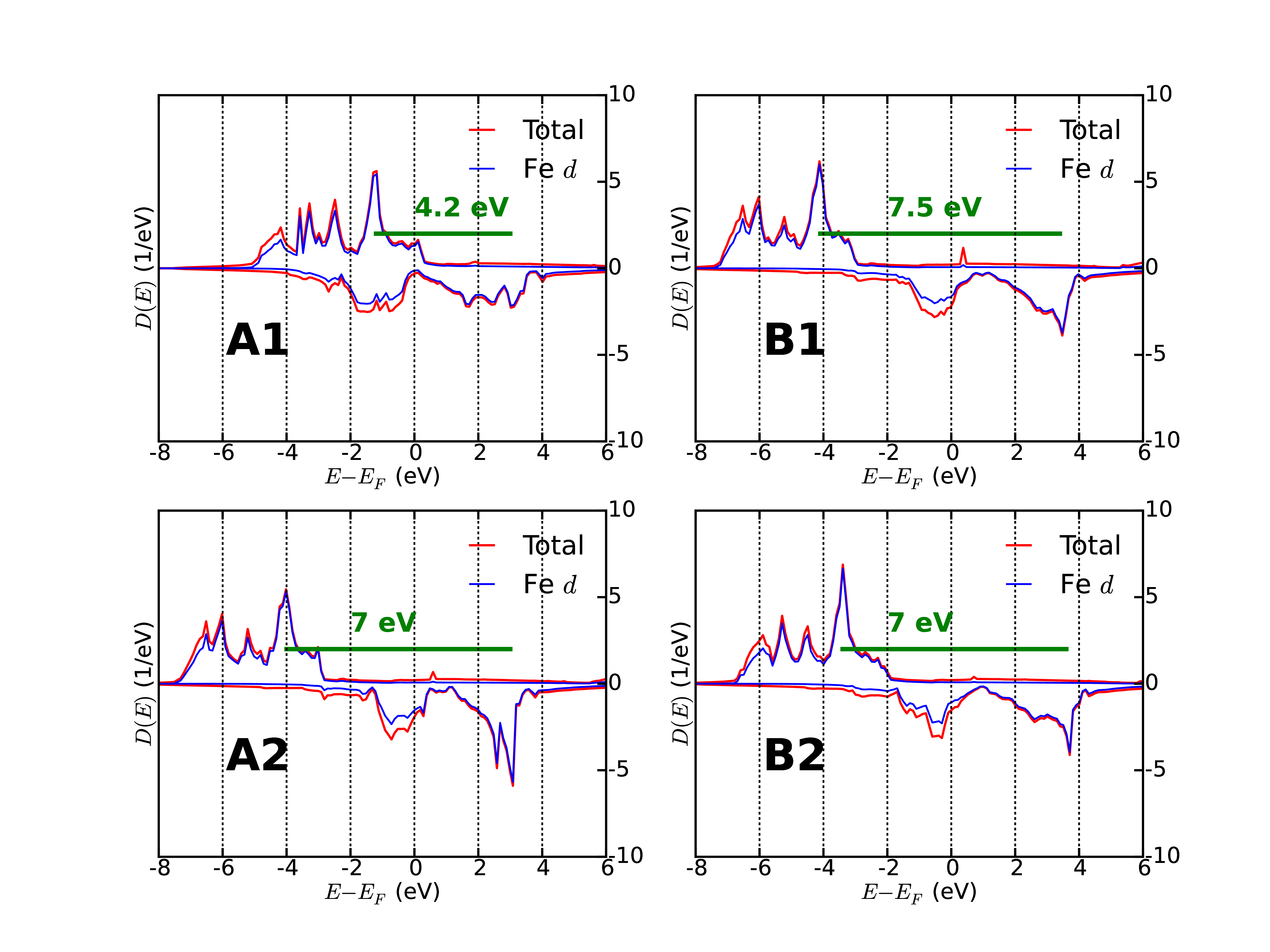}
\caption{\label{fig:fedos} Density of states of body-centered iron with 
ferromagnetic ordering, calculated using different exchange correlation 
functional approximations: \textbf{A1}) LDA+$U$; \textbf{A2}) LDA+$U$+$J$; 
\textbf{B1}) LSDA+$U$; \textbf{B2}) LSDA+$U$+$J$; $U$ = 6 eV and $J$ = 1 eV. 
Horizontal green lines and numbers provide estimates of spin splitting.}
\end{figure}

The spin splitting is defined as in the previous case and is again 
shown as a horizontal bar (green on-line) in
Fig.~\ref{fig:dos}. Comparison of Fig.~\ref{fig:dos}\textbf{A1}
(LDA+$U$) and \textbf{A2} (LDA+$U$+$J$) shows that adding a $J$ to
LDA+$U$ calculations increases the spin splitting by 2.5 eV, slightly
larger than $2J$. The difference arises from a small occupancy of $e_g$
states. 

Comparison of Fig.~\ref{fig:dos}\textbf{A1} (LDA+$U$) and \textbf{A2}
(LDA+$U$+$J$) to \textbf{B1} (LSDA+$U$) reveals that even with no
added $J$, the LSDA+$U$ method produces a larger spin splitting than the
LDA+$U$+$J$ method with $J$ = 1 eV: in other words, the spin
dependence of the exchange correlation functional corresponds to an
effective $J\gtrsim 1$ eV on the transition metal $d$ orbitals. This 
is consistent with the estimate of intrinsic ``$J$'' from pure LSDA 
spectrum (Fig.~\ref{fig:smofedos}).

Inspection of Fig.~\ref{fig:dos}\textbf{B2} (LSDA+$U$+$J$) reveals
that adding a $J$ to the LSDA+$U$ \textit{reduces} the spin splitting, in
contrast to the effect of adding a $J$ to the LDA+$U$ calculation. This 
is consistent with the analysis of our simple $t_{2g}$ model. We 
believe this counterintuitive $J$ dependence in sDFT+$U$+$J$ method 
is a general feature, as was previously noted in the study of a 
spin-crossover molecule~\cite{ChenJia-PRB-2015}. The underlying origin is that
the spin dependence of the double counting correction overcompensates for 
the Hartree shift produced by the $J$, which is consistent with the trend that 
increasing $J$ in sDFT+$U$+$J$ \textit{destabilizes} ferromagnetism.

Fig.~\ref{fig:fedos} presents the density of states of body-centered Fe.
Qualitatively, the variation of spin-splittings predicted by different
exchange correlation functionals (LDA+$U$ $\to$ LDA+$U$+$J$ $\to$
LSDA+$U$ $\to$ LSDA+$U$+$J$) is very similar to that found for cubic
ferromagnetic SrMnO$_3$.

\section{Conclusions \label{sec:conclusion}}

In this paper we have studied energetics and local magnetic moments of
representative transition metal oxides (antiferromagnetic Mott
insulator SrMnO$_3$, moderately correlated metal SrVO$_3$ and negative
charge transfer insulator LaNiO$_3$) and an elemental transition metal
(Fe) to gain further insight into the physics of spin-dependent
density functional theories and their ``+$U$'' and ``+$J$'' extensions
previously noted in Ref.~\cite{Park-PRB-2015}. In these materials, the
only states with significant spin polarization are the transition
metal $d$-states and important aspects of the physics are controlled
by an exchange splitting of spin configurations of these states. For a
transition metal ion in free space, the exchange splitting is
conventionally described by a Hund's coupling parameter $J$ and we
interpret the exchange splitting found in our calculations as an
effective $J_{eff}$, which may have contributions from the spin
dependence of the density functional and from an explicitly added
interaction term.

The results are similar for all materials studied. The spin-dependent
density functionals are found to encode an exchange splitting in the
spin configurations of transition metal $d$-orbitals, which is
larger in the spin-dependent PBE functional (sPBE) than that in the local
spin density functional (LSDA) but in both cases is at least 2 eV. Comparison
to results of the ``+$U$'' ``+$J$'' methods suggests that the
$J_{eff}$ corresponding to the spin-dependent density functional is about 
1 eV. This value is larger than the range of 0.6-1 eV which is
generally accepted as a reasonable estimation for transition metals and 
their oxides, suggesting that the present implementations of the
spin-polarized DFT methods may overestimate the effects of spin
polarization in transition metal-$d$ orbitals.

We also found that including an explicit Hund's coupling $J$ to the
spin dependent DFT functional (sDFT+$U$+$J$) reduces the calculated
exchange splitting below its $J$ = 0 value, whereas adding a $J$ to
the charge-density-only DFT functional (DFT+$U$+$J$) increases the splitting
as expected. This counterintuitive $J$ dependence in sDFT+$U$+$J$
method arises from the spin-dependence of the double counting
correction. The effect was previously noted in the study of
LaNiO$_3$~\cite{Park-PRB-2015} and was carefully documented in the
study of a spin crossover molecule~\cite{ChenJia-PRB-2015}. Our
results provide further support for the previous conclusions
~\cite{Park-PRB-2015, ChenJia-PRB-2015} that while spin-dependent
density functionals provide successful descriptions of many materials,
caution is needed in their applications to transition metals and their
oxides. In particular, for these compounds it is advantageous to base
beyond density functional analyses such as the +$U$+$J$ and +DMFT on
spin-independent density functionals (LDA or the PBE-parametrized
GGA functional), because the physical meaning of $U$ and $J$ in
the parametrization is more clear and the value of $J$ implicit in
present implementations of the spin-dependent exchange correlation
functionals is likely to be too large.

\appendix

\section{Derivation of Eqs.~(\ref{Vup}, \ref{Vdown})}

In this section, we derive Eqs.~(\ref{Vup}, \ref{Vdown}). The Hamiltonian of 
a rotationally invariant Slater-Kanamori (SK) interaction is:

\begin{equation}
\label{eqn:SK} \hat{H}_{\textrm{SK}} = \sum_{\alpha}U\hat{n}_{\alpha\uparrow}
\hat{n}_{\alpha\downarrow}+\frac{1}{2}\sum_{\alpha\neq\beta}(U-2J)
\hat{n}_{\alpha\uparrow}\hat{n}_{\beta\downarrow}
+\frac{1}{2}\sum_{\alpha\neq\beta, \sigma}(U-3J)\hat{n}_{\alpha\sigma}
\hat{n}_{\beta\sigma}
\end{equation}
where $\alpha$ labels a $d$ orbital and $\sigma$ labels a spin. On the 
mean-field level, we simply approximate the operator 
$\hat{n}_{\alpha\sigma}$ as an occupancy $n_{\alpha\sigma}$ and then obtain 
an energy functional:

\begin{equation}
\label{eqn:E} E = \sum_{\alpha}Un_{\alpha\uparrow}
n_{\alpha\downarrow}+\frac{1}{2}\sum_{\alpha\neq\beta}(U-2J)
n_{\alpha\uparrow}n_{\beta\downarrow}
+\frac{1}{2}\sum_{\alpha\neq\beta, \sigma}(U-3J)n_{\alpha\sigma}
n_{\beta\sigma}
\end{equation}
The potential associated with a given orbital $\alpha$ and a given 
spin $\sigma$ is $V^{\sigma}_{\alpha}= \frac{\partial E}{\partial n_{\alpha\sigma}}$.
Therefore we have:

\begin{eqnarray}
V^{\uparrow}_{\alpha} = \frac{\partial E}{\partial n_{\alpha\uparrow}}=
Un_{\alpha\downarrow}+\sum_{\beta\neq\alpha}(U-2J)n_{\beta\downarrow}
+\sum_{\beta\neq\alpha}(U-3J)n_{\beta\uparrow}\label{eqn:upV}\\
V^{\downarrow}_{\alpha} = \frac{\partial E}{\partial n_{\alpha\downarrow}}=
Un_{\alpha\uparrow}+\sum_{\beta\neq\alpha}(U-2J)n_{\beta\uparrow}
+\sum_{\beta\neq\alpha}(U-3J)n_{\beta\downarrow}\label{eqn:downV}
\end{eqnarray}
For the simple model of a half-filled fully spin-polarized
orbitally symmetric $t_{2g}$ shell, we have three orbitals: 
$\alpha = d_{xy}, d_{xz}, d_{yz}$ and the occupancy is: $n_{\alpha\uparrow}=1$, 
$n_{\alpha\downarrow}=0$ for each orbital $\alpha$. Therefore we have:

\begin{eqnarray}
V^{\uparrow}_{\alpha} = 0+0+2(U-3J)=2U-6J\label{eqn:upVV}\\
V^{\downarrow}_{\alpha} =U+0+2(U-2J)=3U-4J\label{eqn:downVV}
\end{eqnarray}
which are Eqs.~(\ref{Vup}, \ref{Vdown}).

\begin{acknowledgments}
A. J. Millis is supported by National Science Foundation under grant
No. DMR-1308236. H. Chen is supported by National Science Foundation
under Grant No. DMR-1120296. A. J. Millis thanks the College de France
for hospitality and a stimulating intellectual environment while this
paper was being prepared. We thank C. Marianetti, A. Georges,
S. Biermann and especially, K. Burke for helpful comments.
\end{acknowledgments}

\bibliography{xc-functional-v6}

\end{document}